\documentstyle[aps,prl,preprint]{revtex}
\begin{document}
\draft

\title{Coordinate Conditions and Their Implementation \\ 
              in 3D Numerical Relativity }
\author{Jayashree Balakrishna$^1$, Gregory Daues$^1$, Edward Seidel$^2$, Wai-Mo Suen$^{1,2}$, Malcolm Tobias$^1$,
and Edward Wang$^1$}

\address{${}^1$McDonnell Center for the Space Sciences, Department of Physics,
Washington University, St. Louis, Missouri 63130}

\address{${}^2$National Center for Supercomputing Applications,
605 E. Springfield Ave., Champaign, Illinois 61820}

\date{Submitted 1/11/96}

\maketitle

\begin{abstract}
We put forth a few ideas on coordinate conditions and their
implementation in numerical relativity. Coordinate conditions are
important for the long time scale simulations of relativistic systems,
e.g., for the determination of gravitational 
waveforms from astrophysical events to be measured by LIGO/VIRGO.  We
demonstrate the importance of, and propose methods for, the {\it
active enforcement} of coordinate properties.  In
particular, the constancy of the determinant of the spatial 3-metric
is investigated as such a property.  We propose an exceedingly simple
but powerful idea for implementing
elliptic coordinate conditions that not only makes possible the use
of complicated elliptic conditions, but is also {\it orders of
magnitude} more efficient 
than existing methods for large scale 3D simulations.

\end{abstract}

\pacs{PACS numbers:  04.25.Dm, 04.30.-w, 95.30.Sf}

{\it \underline{Introduction.}} It was suggested that the `Holy
Grail of Numerical Relativity'\cite{teuk} is a 3D code that handles black holes by
avoiding singularities, maintains accuracy, and is capable of running
forever.  We believe, with the apparent horizon boundary condition
(AHBC)\cite{ahbc}, singularities may no longer be an 
important threat.  Instead,
the major difficulty we face, in long time scale evolutions of 3D
spacetimes, is to control the coordinate freedom. 
The aim of this letter is to present a few concepts that
are important for formulating, and implementing, good coordinate conditions.

In a numerical simulation the relative motion of coordinate grid
points is affected by:
(a) gauge choices requiring the metric to
take up certain forms (e.g., a diagonal 3-metric);
(b) shift vectors which shift the coordinate points from moving normal to the
constant time slice; and in particular we want to emphasize,
(c) the time slicing condition, 
which among other things also determines the
normal direction of motion of grid points.

The use of gauge conditions of type (a) above is common in lower-dimensional
studies, where some symmetries of the spacetime
are assumed (e.g., spherical symmetry).
In such cases, there are often preferred choices of the 
form of the 3-metric (e.g., $g_{\theta \theta} = r ^ 2 $). 
However, for a general 3$+$1 spacetime
without any symmetry assumptions, there is {\it a priori}
no preferred choice and we have to make use of the
shift and the lapse to control the 
coordinate degrees of freedom. Both of these
are essential for good control.

Time slicing conditions and shift vectors 
in long time scale 3D evolutions have not
been studied much, as previous work in numerical relativity
has mostly been carried out for lower-dimensional systems
(e.g., axisymmetric systems, where gauge choices in the sense
of (a) are effective).
Time slicing conditions have been studied, mainly as
a tool for avoiding singularities in black hole
spacetimes; shift conditions have
been used with less success, 
except in the form of (a) above, and in
AHBC  work \cite{ahbc}.
As we are now forging ahead into full 3D numerical relativity,
aiming at long stable evolutions, the choice of the coordinate
conditions is crucial.

The long term numerical evolution of full 3D systems is an
important goal for numerical relativity, especially for 
determining the gravitational radiation from 
realistic astrophysical events that LIGO/VIRGO may detect.
For example, for radiation from inspiraling compact binary systems
(one of the most important sources expected for LIGO/VIRGO), one
would like to be able to numerically evolve the binary system
for hundreds or even thousands of $M$, where $M$ is the mass of 
the system. 

{\it \underline{Underlying Ideas.}}
In this paper we study the coordinate conditions in the 3$+$1
formulation in Cartesian coordinates.
The ideas underlying the present study in controlling the lapse $
\alpha $ and the shift $ \beta^i $ are the following:

\noindent
(i) To distinguish good from bad coordinate conditions, one
important criterion is: Any {\it secular changes } in the metric
should be due solely to the secular changes of the spacetime geometry,
not the gauge degrees of freedom.  In this paper, by stable evolutions we
mean not just numerical
stability but also the absence of ``secular drifting'' due to gauge freedom. 

\noindent
(ii)   In order to obtain
a stable evolution over a long time scale, it is important
to ensure that the coordinate conditions used are not only
suitable for the geometry of the spacetime being evolved, but also
that {\it the conditions themselves are stable}. That is, when the 
condition is perturbed, e.g., by an inaccuracy 
in the numerical evolution, there is no long term secular drifting. 

\noindent
(iii) Instead of the lapse,
what is more important is the {\it lapse history}. 

\noindent
(iv) The lapse and shift are interrelated.  
They should not be considered independently.

\noindent
(v) An exact 
implementation of a coordinate condition is unnecessary. 
An approximate but stable implementation can lead
to an {\it accurate} construction of the spacetime. 

{\it \underline{Driver Conditions.} } Here we investigate the {\it
active enforcement} of a coordinate property. The emphasis 
here is {\it not} on what properties of the shift or lapse one
wants to enforce, but rather the need, and how, to actively enforce a
property in 3D numerical evolutions. We begin with a familiar case.
The maximal slicing condition\cite{york}
\begin{equation}
K = 0 = \frac{ \partial K }{ \partial t } ,
\label{k=0}
\end{equation}
leads to the elliptic equation for the lapse 
(assuming the Hamiltonian constraint),
\begin{equation}
   \nabla^2 \alpha - K_{ij} K^{ij} \alpha  
- 4 \pi ( S + \rho ) \alpha = 0  , 
\label{maxsl}
\end{equation}
where $ \rho $ is the energy density and $ S $ is the trace of 
the spatial stress tensor. 

To demonstrate the {\it deficiency} of (\ref{maxsl}) 
as a coordinate condition,
we first study an evolution 
of flat space with zero shift and an initial lapse which is close to 
but not exactly unity
\begin{equation}
\alpha ( t = 0 ) = 1.0 + 0.001 \, \exp [ - (x^2 + y^2 + z^2) ] .
\end{equation}
We evolve the initial data $g_{ij}= \delta _{ij}$ and $K_{ij}=0$
forward with this lapse for just the first timestep, with
subsequent lapses determined by (\ref{maxsl}).  The code
used in this study has been described in \cite{3dbh}.  The lapse at
various times is given in Fig.\ 1(a). 
We see that the lapse
returns to nearly unity after the first step, as it should. 
One might expect a healthy evolution.
To see what actually happens, in Fig.\ 1(b) we show the
evolution of a typical metric function $ g_{xx} $ along the
$ z $ axis. 
The dashed lines give $ g_{xx} $ up to $ t = 8 $ 
at intervals $ \Delta t = 2 $ under the  
maximal slicing condition (\ref{maxsl}).  
We see that a dip is developing. Eventually this feature is too
sharp to be resolved and the code crashes.

The reason behind this development is clear. The maximal slicing
Eq.~(\ref{maxsl}) {\it
does not actively 
enforce} the desired condition $ K = 0 $.  That is, if $\alpha$ is perturbed 
at any time, making $ K $ nonzero, the slicing condition
(\ref{maxsl}) cannot put $ K $ back to zero. In Fig.\ 1(c) the dashed
lines show the evolution of $K$. With the lapse nearly returned to one but
the slicing still ``curved'' (with respect to Minkowski coordinates),
this ``curvature'' will be preserved by the evolution. The grid points moving
orthogonal to the slicing drift with respect to one another, leading
to secular evolution of the metric functions.  This example gives a
simple yet vivid demonstration that what is important is not only
the lapse but the {\it history} of the lapse.  Although for
most of the evolution the lapse is nearly one, the preferred value for flat
spacetime, the slicing is bad as errors were made in the
past. The value of the lapse at any instant in time does not 
tell us much about the ``shape'' of the constant time slice.  One
needs to control the history.

In this flat space study, the perturbation of the lapse at $ t = 0 $
is put in by hand. In the actual evolution of 3D relativistic systems
involving highly nonlinear and dynamical fields, it is difficult
if at all possible to ensure that any coordinate condition is
precisely kept everywhere and for all time.  Once the condition is perturbed
by numerical inaccuracy, without active enforcement,
the phenomena shown here can appear. 

In the following 
we show the numerical evolution of the metric functions
for an equilibrium compact stellar configuration 
made up of a self-gravitating scalar field \cite{boson}.
As the initial configuration has $K=0$, and is in equilibrium,
the metric should remain constant in time with maximal slicing.
The actual numerical evolution
of $ g_{rr} $ (calculated from the Cartesian
metric functions $ g_{xx} $, $ g_{xy} $, etc.\ 
evolved in the code \cite{3dbh}) for maximal slicing is shown 
with dashed lines at various times in Fig.\ 1(d). 
We see a secular drift, which is caused
by $ K $ drifting away from zero due to 
numerical inaccuracy.  There is no control over this drifting with 
maximal slicing.

For all of these cases, we see
that the secular drift is related to the fact that Eq.
(\ref{maxsl}) does not actively enforce Eq.\ (\ref{k=0}).
Can one construct a slicing condition which 
{\it drives} $ K $ back to zero 
when it is perturbed away
from it? This leads to what we call the ``$ K $-driver''
slicing condition
\begin{equation}
  \frac{ \partial K }{ \partial t } + c K = 0 , 
       \label{kdr}
\end{equation}
where $ c $ is a positive number of our choice. If $ c $
is a constant in time, $ K $ is driven to zero in an 
exponential manner. By controlling $ c $, one has control 
over {\it the stability of the slicing}. Eq.\ (\ref{kdr}) leads to the
elliptic equation for the lapse
\begin{equation}
 \nabla^2 \alpha - K_{ij} K^{ij} \alpha - \beta ^i \nabla _ i K 
   - 4 \pi ( S + \rho ) \alpha  = c K  .
\label{kdr2}  
\end{equation}
The working of the $ K $-driver slicing is shown in Fig.\ 1 
as the solid lines. 
We see that in each case (Fig.\ 1(a-d)) the secular drift is under control.
There is no assumption of $ K = 0 $ made, but $ K $ is held 
much closer to zero 
with (\ref{kdr2}) than (\ref{maxsl}) as shown in Fig.\ 1(c) \cite{eppley}.

We emphasize that the idea of a ``driver'' is more general
than the particular case of (\ref{kdr2}), which is just one 
possible way of
actively enforcing a coordinate property. 
Other ways of controlling $ K $  are possible, 
as will be analyzed below, and elsewhere \cite{follow} .

The idea of a driver can be applied to other coordinate
properties.
We next show what we call the  $\gamma$-driver for the lapse, 
based on the condition~\cite{some}
\begin{equation}
\partial_t ( \log \sqrt{ \gamma} ) = - \alpha K 
                       + \nabla_i \beta^i  =  0 ,
\label{gdr}
\end{equation}
where $ \gamma = \det(\gamma_{ij})$. 
To actively enforce this ``constant $ \gamma $'' condition, we
use the ``$\gamma$-driver slicing'' 
\begin{equation}
 \frac{ \partial }{ \partial t } 
 \left( - K + \frac{ \nabla_i \beta^i }{ \alpha }  \right)
   = - c \,  \left( - K + \frac{ \nabla_i \beta^i }{ \alpha }  \right)  ,
\label{gdr2}
\end{equation}
leading to the elliptic equation for the lapse
\begin{eqnarray}
 &\,& \nabla^2 \alpha - K_{ij} K^{ij} \alpha - \beta ^i \nabla _ i K
       - 4 \pi ( S + \rho ) \alpha   \nonumber   \\ 
 &\,&     \qquad\qquad\qquad    =  - \frac{ \partial }{ \partial t } 
     \left(  \frac{ \nabla_i \beta^i }{ \alpha } \right) 
   + c \left( K - \frac{ \nabla_i \beta^i }{ \alpha } \right)  , 
\label{gdr3}
\end{eqnarray} 
where the right-hand side is treated as a source term to be evaluated
on the previous slice.

We note that (\ref{gdr3}) is uncommon as a slicing condition, 
as it explicitly involves the shift. Traditionally, when
considering slicing conditions, the attention is on obtaining
a good foliation of the spacetime (hence shift independent).  
However, we believe that due to 
the effect of the lapse on the motion of the
coordinate grid, foliation is not the only concern.
Notice that when the shift is zero, (\ref{gdr}) reduces to 
$ K = 0 $, the maximal slicing.  
For a nonzero shift whose divergence
has nonzero time-average (e.g., in AHBC \cite{ahbc}),
maximal slicing leads to a secular change in $ \gamma $, cf.\ (\ref{gdr}). 
The constant $ \gamma $ slicing of (\ref{gdr3}) 
does not have this pathology.

The $\gamma$-driver slicing (\ref{gdr3}) is not singularity avoiding,
depending on the shift.  This is not a drawback, when used with our
``AHBC'' scheme~\cite{ahbc} for handling singularities.  In Fig.\ 2 the
solid line gives the evolution of the determinant of the 3-metric
$\gamma$, for the case of a spherical black hole, obtained using
(\ref{gdr3}) with our ``AHBC'' code~\cite{ahbc}.  The shift used is
the area-locking shift~\cite{ahbc}.  This is to be compared to the
dashed line, obtained using the same grid parameters but with the $ K
$-driver (\ref{kdr2}).  The advantage of the $ \gamma $-driver in
terms of stability against secular drifting of the metric is apparent.
Further details of the comparison and the applications of (\ref{gdr3})
to the evolution of neutron stars and gravitational wave packets will
be given in a follow up paper \cite{follow}.

{\it \underline{Elliptic conditions without elliptic
equations.}} In 3D numerical relativity elliptic coordinate
conditions (e.g., maximal slicing, or the $ K $ and $ \gamma $-drivers
studied above) are often preferred, as they can provide global
smoothness.  However, elliptic equations are computationally
expensive, particularly in 3D with large numbers of grid points.
Nonlinear elliptic equations can be complicated and algorithms
suitable for solving them are yet to be developed, e.g., for the 3-coupled
minimal distortion equations in 3D.  Furthermore, complicated boundary
conditions, e.g., an irregular shaped inner boundary in the case of
determining the shift or lapse with an AHBC, may make a global
solution to the elliptic equation very difficult if at all possible.

All these problems can be solved in one stroke,
based on the idea that an approximate implementation of
the coordinate condition is good enough as long as it is stable.
The spacetime constructed can still in principle be exact
even when the implementation of the coordinate conditions 
is approximate.

The idea is to ``evolve the elliptic equation''
by rewriting it in a parabolic form. For example,
instead of directly solving the elliptic equation
(\ref{kdr2}), we take 
\begin{equation}
\frac{\partial \alpha }{ \partial t } 
 =  \epsilon \, (  \nabla^2 \alpha - K_{ij} K^{ij} \alpha
- \beta ^i \nabla _ i K - 4 \pi ( S + \rho ) \alpha - c K ) . 
\label{evolm} 
\end{equation}
This is nothing more than the Richardson method
\cite{saylor} of
solving elliptic equations, but carried out {\it in real time
as part of the evolution}.  As long as $ dt $ 
(the timestep in finite differencing) 
is much smaller than 
(i) the dynamical time scales of the spacetime and 
(ii) the instability timescale (if any) due to the small 
violation of the coordinate condition, 
there are many timesteps available for 
(\ref{evolm}) to relax to (\ref{kdr2}).  Note that for implementation
of coordinate conditions, the goal is {\it not} to precisely maintain the  
condition, but to provide a stable evolution.  

Fig.\ 3(a) shows the evolution of $g_{zz}$ for a 3D Brill wave evolved
with the lapse given by (\ref{evolm}) and the shift by
\begin{eqnarray}
\frac{\partial \beta_{i} }{ \partial t } 
 &=& \epsilon \, \left( \nabla^2 \beta_{i} +
        \frac{1}{3} \nabla_{i} ( \nabla^{j} \beta_{j} ) +
        R^{j}_{i} \beta_{j}\;  -      \right. \nonumber  \\   
    &\,&  \qquad\qquad\qquad \left.
        \nabla ^{j} \left[ 2 \alpha(K_{ij} - \frac{1}{3} K g_{ij})
        \right] \right) ,
\label{mini} 
\end{eqnarray}
which is the minimal distortion shift \cite{york},
but now implemented through evolution.  
To the best of our knowledge, this is the
first time that the full minimal distortion shift can be
implemented in 3D. The result can be compared to the 
dashed lines obtained
using the traditional maximal slicing equation
(\ref{maxsl}) with zero shift.  We see that the evolution represented
by the solid lines is more stable than that of the dashed lines at late times. 
The point we want to make here is not
really the comparison of stability, but rather that a fully coupled, 
complicated set of lapse and shift conditions can be implemented 
with this ``evolving elliptic equation'' method.

Two technical points are worth mentioning.
First, the optimum choice for the value of $ \epsilon $ 
is nontrivial \cite{saylor} and depends on the various time scales involved 
in the numerical evolution. Second, there are situations
where it is desirable to iterate more than once
on a single time slice to achieve a more thorough 
relaxation. As the metric functions are regarded as given on
each time slice, the added computational expense 
is not significant \cite{follow}.

Recently we have extensively used this ``evolving elliptic
equations'' scheme in implementing coordinate conditions, for our
studies of black holes and gravitational waves.  The improvement in
efficiency, without sacrificing accuracy or stability, can be
dramatic, especially when large grids are used.
In Fig.\ 3b the CPU time spent solving maximal slicing (\ref{maxsl})
using CMSTAB \cite{3dbh}(boxes) versus evolving (\ref{evolm}) (pluses),
is shown for various grid sizes for the Brill wave evolution.   

This evolving scheme also makes possible the use of
elliptic conditions, e.g., the minimal distortion shift, for
black hole studies with a dynamically evolving inner boundary (AHBC).
The details of these applications will 
be reported elsewhere \cite{follow}. 
We thank Joan Mass\'o and Paul Saylor for useful discussions.
The calculations in this paper were performed with the CM5 at NCSA and the
C90 at PSC.
This research is supported by PHY94-04788, PHY94-07882.

%%%%%%%%%%%%%%%%%%%%%%%%
%%%%%%%%%bibliography%%%%%%%%%%%%%%
%\bibliographystyle{prsty}
%\bibliography{/afs/.ncsa.uiuc.edu/projects/genrel/share/Texfiles/Bibtex/references}
% \bibliography{references}
%%%%%%%%%%%%%%%%%%%%%%%%

%%%%%%%%%Figures%%%%%%%%%%%%%%

%1
\begin{figure}
\caption{(a) The lapse at various times is shown.  The initial ($t=0$)
lapse is put in by hand.  Subsequently we solve either the maximal
slicing Eq.\ (2)
(dashed lines) or the $ K $-driver Eq.\ (5) with $c={1}/ ({5dt}) $
(solid lines)  
for the lapse.  In either case they return to one quickly as expected.
(b) $g_{xx}$ at various times is shown.  The maximal
slicing result shows a secular drift, while the $K$-driver result is 
stable.
(c) The value of $ K $ reaches a non-zero profile with maximal slicing,
while the $K$-driver actively enforces $ K=0 $.
(d) Here we compare maximal slicing to the $K$-driver
for the numerical evolution of a compact self-gravitating
scalar field in equilibrium.  In this case, the lapse is perturbed from
the exact maximal slicing value only by
numerical inaccuracy.
}
\label{fig:fig1}
\end{figure}

\begin{figure}
\caption{The evolutions of the determinant of the 3-metric $\gamma$
for a spherical black hole in the ``AHBC'' code are compared
for the cases of $K$-driver (dashed lines) vs.\ $\gamma$-driver
(solid lines).  The runs are done with the same grid parameters.
The secular drift is negligible with $\gamma$-driver at late time.
}
\label{fig:fig2}
\end{figure}

\begin{figure}
\caption{(a) The evolution of the metric component $g_{zz}$ is shown
for a Brill wave.  The evolution is obtained with maximal slicing Eq.\ (2)
(dashed lines) and by evolving the $K$-driver Eq.\ (9) 
with the minimal distortion
shift Eq.\ (10) (solid lines).  $\epsilon = 0.1 {dx ^ 2} / {dt}$ and
$c={1} / (20dt)$ are  
used. (b) We compare the computational cost of
integrating an elliptic lapse condition (boxes) against evolving
(pluses) for various sized grids.
}
\label{fig:fig3}
\end{figure}

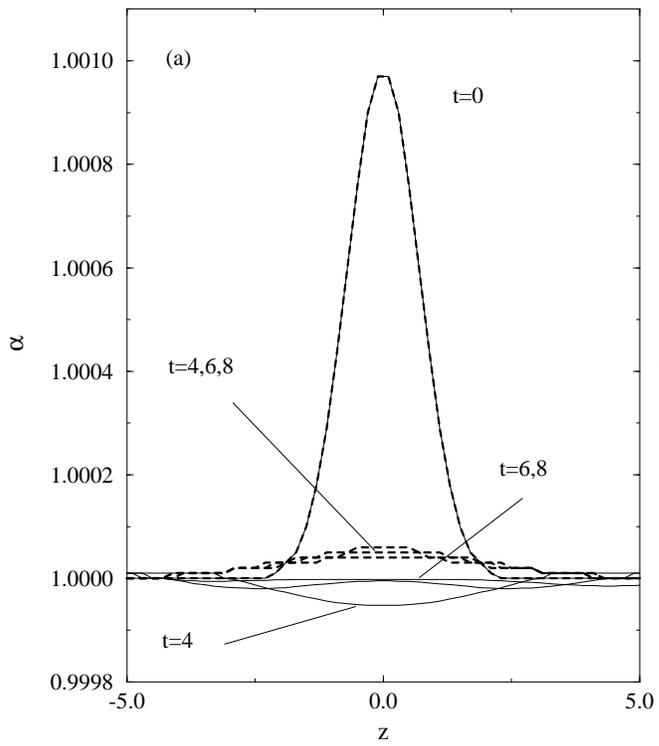
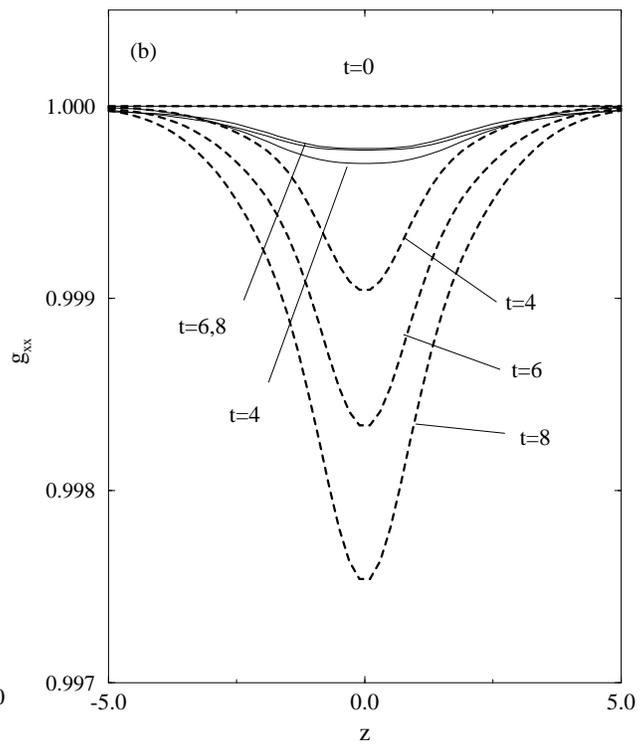
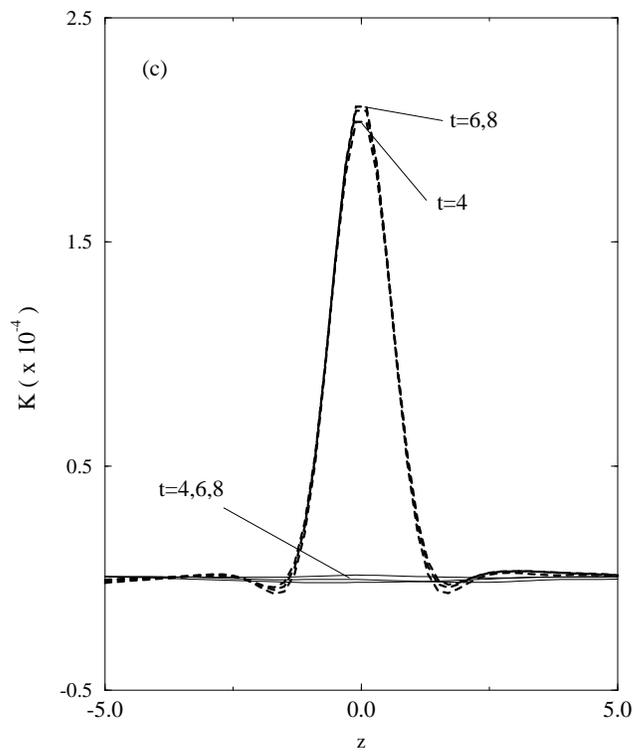
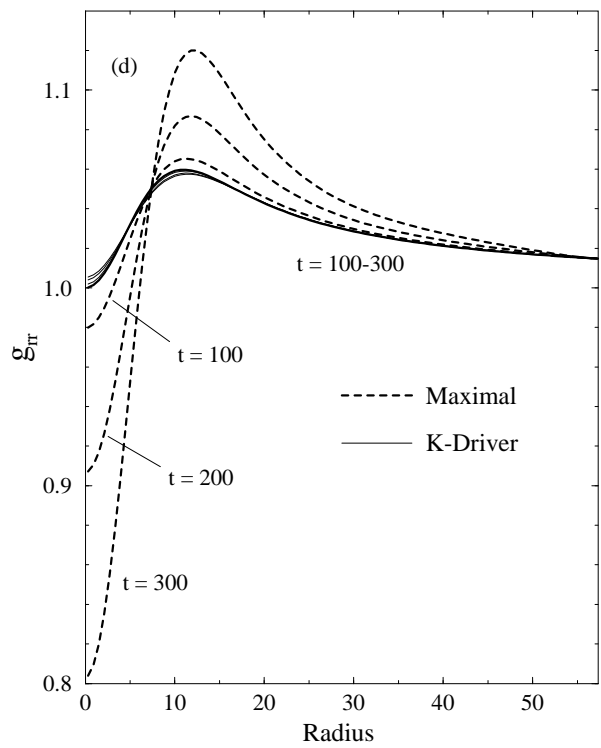
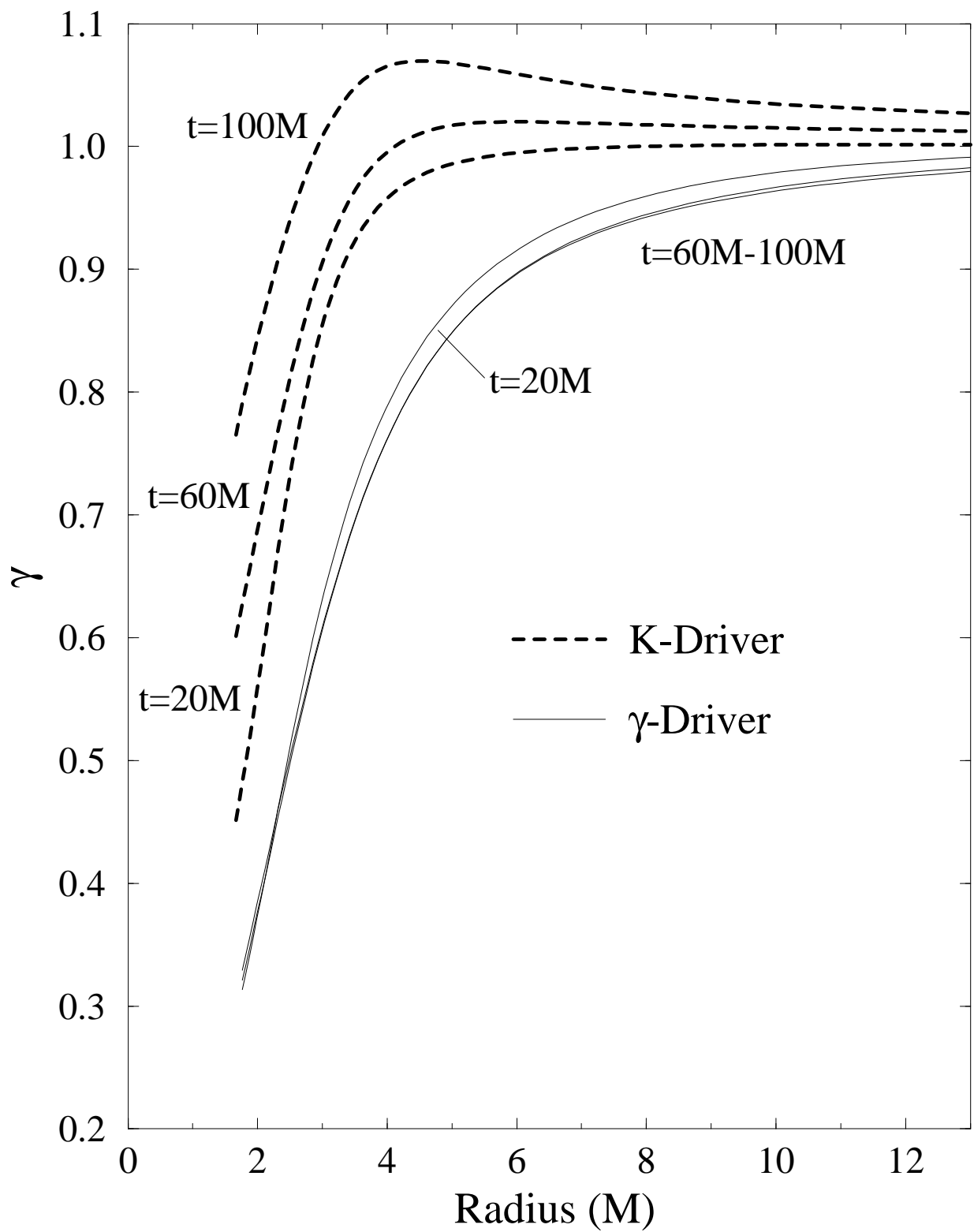
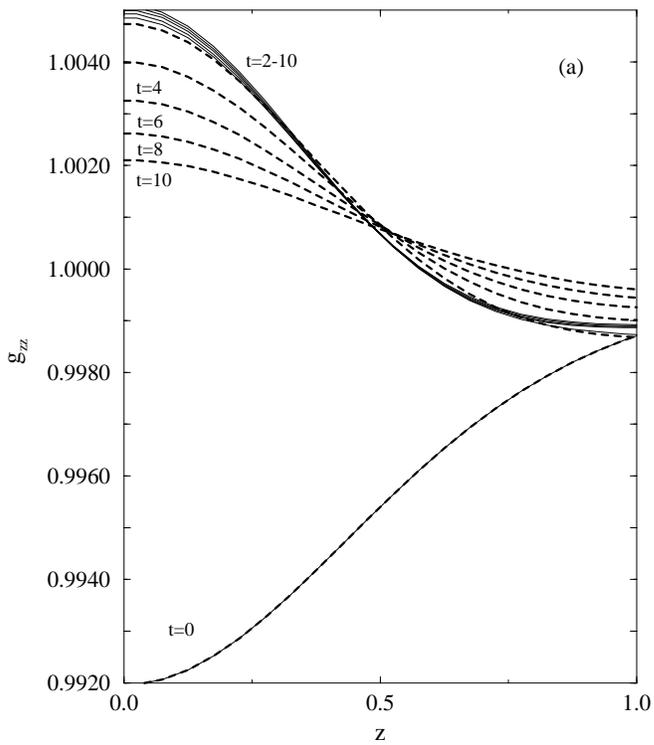
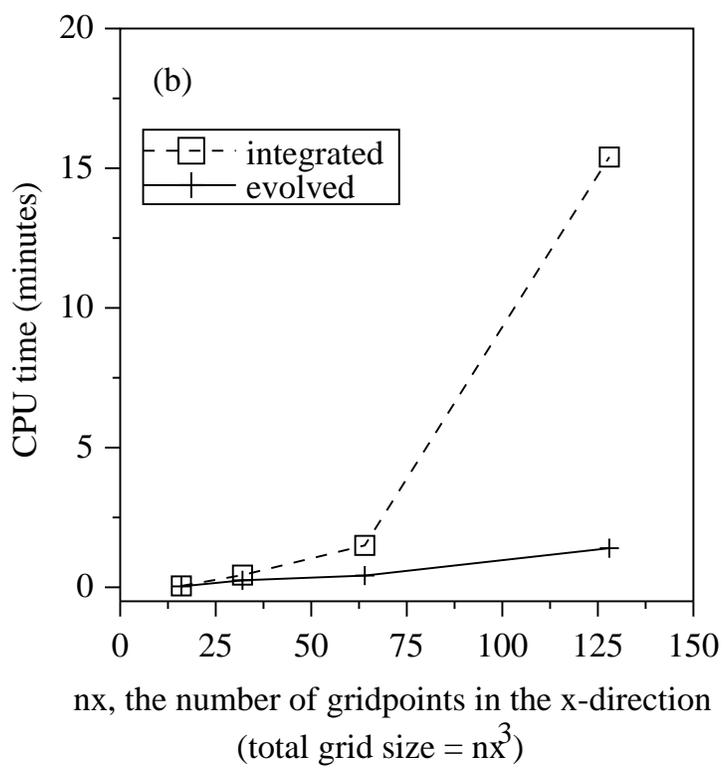
\end{document}